\documentclass[a4paper]{jpconf}
\usepackage{amsmath}
\usepackage{amssymb}
\usepackage{graphicx}
\usepackage{bm}
\begin{document}
\title{Vortex Core Structure in Multilayered Rashba Superconductors}

\author{Y Higashi$^{1}$, Y Nagai$^{2}$, T Yoshida$^{3}$ and Y Yanase$^{4}$}

\address{$^{1}$Department of Mathematical Sciences, Osaka Prefecture University, 1-1 Gakuen-cho, Sakai 599-8531, Japan}
\address{$^{2}$CCSE, Japan Atomic Energy Agency, 5-1-5 Kashiwanoha, Kashiwa, Chiba 277-8587, Japan}
\address{$^{3}$Graduate School of Science and Technology, Niigata University, Niigata 950-2181, Japan}
\address{$^{4}$Department of Physics, Niigata University, Niigata 950-2181, Japan}

\ead{higashiyoichi@ms.osakafu-u.ac.jp}
\begin{abstract}
We numerically study the electronic structure of a single vortex in two dimensional superconducting bilayer systems within the range of the mean-field theory.
The lack of local inversion symmetry in the system is taken into account through the layer dependent Rashba spin-orbit coupling.
The spatial profiles of the pair potential and the local quasiparticle density of states are calculated in the clean spin-singlet superconductor on the basis of the quasiclassical theory.
In particular, we discuss the characteristic core structure in the pair-density wave state,
which is spatially modulated exotic superconducting phase in a high magnetic field. 
\end{abstract}
\section{Introduction}
Intensive studies on locally noncentrosymmetric systems (LNCS) has been conducted \cite{yoshida2014,sigrist2014}.
The LNCS are realized in various materials such as multilayered cuprates \cite{mukuda2012}, the doped topological insulator Cu$_x$Bi$_2$Se$_3$ \cite{hor2010}, the pnictide superconductor SrPtAs \cite{nishikubo2011} and so on.
Especially, the heavy-fermion CeCoIn$_5$/YbCoIn$_5$ superlattice is fabricated \cite{mizukami2011} and the artificial control of the inversion symmetry breaking was realized recently \cite{shimozawa2014}.
Therefore, one can seek experimentally the theoretically predicted exotic superconducting phase in a magnetic field \cite{yoshida2012,yoshida2013}.
The inhomogeneity due to a magnetic field is not taken into account in the existing theoretical researches on the superconductivity in LNCS \cite{yoshida2012,yoshida2013}.
However, the exotic superconducting phase stabilizes in a magnetic field.
So understanding the vortex states in LNCS is necessary.
The observation of quasiparticle states in the vicinity of a vortex gives us a lot of information on the exotic superconducting phase.
We focus on the exotic superconducting phase called the pair-density wave (PDW) phase,
which is stabilized in a high magnetic field perpendicular to the layers \cite{yoshida2012}.
In the PDW phase,
the phase of the superconducting order parameter changes by $\pi$ between the two layers,
whereas no phase difference of the order parameter in the BCS phase in a low magnetic field.
In this paper,
we investigate the vortex core structure of the PDW and the BCS phases,
and try to obtain the characteristics of the PDW phase.

\section{Formulation}
We consider the multilayered system characterized by the layer-dependent spin-orbit coupling (SOC) strength $\alpha_m$ with the layer index $m$.
In this paper, the number of the layer is fixed to $N=2$.
The layer dependence of the SOC $(\alpha_1,\alpha_2)=(\alpha,-\alpha)$ reflects the local noncentrosymmetricity of the system \cite{maruyama2012}.
We assume the simple form of the Rashba type SOC described by $\bm{g}(\bm{k})=(-k_y,k_x,0)/k_{\rm F}$, which satisfies the normalization condition $\langle \bm{g}(\bm{k}) \rangle_{\bm{k}}=1$ on the two dimensional Fermi surface (FS).
$k_{\rm F}$ is the Fermi wave number and $\langle \cdots \rangle_{\bm{k}}$ is the average over the FS.

We consider spin-singlet $s$-wave pairing states within the layer.
The pairing potential in the spin space is expressed as $\varDelta_{s_1 s_2}(\bm{r})=\varDelta(\bm{r})(i\sigma_y)_{s_1 s_2}$.
We assume that the spatial variation of the pair potential $\varDelta(\bm{r})$ is the same in each layer.
Then, the pair potential is expressed as $\hat{\varDelta}(\bm{r})=\varDelta(\bm{r}) i\sigma_y \otimes D$,
where $D$ is the $N \times N$ diagonal matrix in the space composed of the layer (band) degree of freedom.
In the case of $N=2$, $D={\rm diag}(1,s)$ with $s=1$ for the BCS state and $s=-1$ for the PDW one.
$\hat{\cdot}$ denotes the $2N \times 2N$ matrix in the spin and band space.
A vortex line perpendicular to the two dimensional superconducting layer has the form,
$\varDelta(\bm{r})=|\varDelta(r)|{\rm exp}(i\phi_r)$ in each layer.

In this paper,
we consider the superconductor in which the paramagnetic pair breaking effect is dominant.
In addition,
we assume the system is in the type-II limit and neglect the vector potential.
Then, we incorporate only the paramagnetic pair breaking effect into the quasiclassical theory through the Zeeman term.
There is three factors which split the 4-fold degenerated FS.
The spin degeneracy is lifted by both the SOC and the Zeeman field $\mu_{\rm B}H$,
and the band degeneracy is lifted by the interlayer hopping $t_\perp$.
We assume that the strength of these three factors is sufficiently weak and incorporate them into the quasiclassical theory as perturbations.
In this situation, the split of the FS is small and
the difference of the Fermi velocity between the weakly split FS is negligibly small.
So we can put $\bm{v}_{\rm F}=v_{\rm F}\tilde{\bm{k}}$.
Then, we can expand the quasiclassical approximation described in Ref.~\cite{hayashi2006} into the multilayered system.

The quasiclassical Green's function depends on the direction $\tilde{\bm{k}}=(\cos \phi_k,\sin \phi_k)$ with the azimuthal angle $\phi_k$ and is written as a $4N \times 4N$ matrix in the particle-hole space:
\begin{equation}
\check{g}(\bm{r},\tilde{\bm{k}},i\omega_n)=-i\pi
\left(
\begin{array}{cc}
\hat{g} & i\hat{f}  \\
-i\hat{\bar{f}} & -\hat{\bar{g}} 
\end{array}
\right),
\end{equation}
where $\omega_n$ is the Matsubara frequency.
The quasiclassical Green's function obeys the following Eilenberger equation with the spin quantization axis parallel to the $z$ axis \cite{seren1983}.
\begin{align}
&i\bm{v}_{\rm F}(\tilde{\bm{k}}) \cdot \bm{\nabla} \check{g}(\bm{r},\tilde{\bm{k}},i\omega_n)
+\left[i\omega_n \check{\tau}_3 - \check{\varDelta}(\bm{r})-\check{K}(\tilde{\bm{k}}),\check{g}(\bm{r},\tilde{\bm{k}},i\omega_n) \right]
= \check{0},\\
\check{K}(\tilde{\bm{k}}) & \equiv
\left(
\begin{array}{cc}
t_{\perp} \hat{H}_{\rm inter} +\mu_{\rm B}H \hat{H}_{\rm Z} & \hat{0} \\
\hat{0} & t_{\perp} \hat{H}_{\rm inter} + \mu_{\rm B} H \hat{H}_{\rm Z}
\end{array}
\right)
+
\alpha
\left(
\begin{array}{cc}
\hat{H}_{\rm SO}(\tilde{\bm{k}}) & \hat{0} \\
\hat{0} & \hat{H}^{\ast}_{\rm SO}(-\tilde{\bm{k}})
\end{array}
\right),
\end{align}
with
\begin{alignat}{1}
\check{\tau}_3
&=
\left(
\begin{array}{cc}
\sigma_0 \otimes I_{N \times N} & \hat{0} \\
\hat{0} & -\sigma_0 \otimes I_{N \times N}
\end{array}
\right),
~~~
\check{\varDelta}(\bm{r})=
\left(
\begin{array}{cc}
\hat{0} & \hat{\varDelta}(\bm{r}) \\
-\hat{\varDelta}^\dag(\bm{r}) & \hat{0}
\end{array}
\right),
\\
\hat{H}_{\rm Z}
&=
-\sigma_z \otimes I_{N \times N},
~~~
\hat{H}_{\rm inter}
=
\sigma_0 \otimes T_{\perp},
~~~
\hat{H}_{\rm SO}(\tilde{\bm{k}})
=
\bm{g}(\tilde{\bm{k}}) \cdot \bm{\sigma} \otimes S_{\rm d}.
\end{alignat}
Here, $I_{N \times N}$ is the unit matrix, $T_\perp={\rm offdiag}(1,1)$ and $S_{\rm d}={\rm diag}(1,-1)$.
$\rm offdiag(\cdot,\cdot)$ indicates the $2 \times 2$ matrix which has the offdiagonal component only.
$\hat{H}^\ast_{\rm SO}(-\tilde{\bm{k}})=-\bm{g}(\tilde{\bm{k}}) \cdot \bm{\sigma}^{\rm tr} \otimes S_{\rm d},$ which is obtained from the relation $\bm{g}(-\tilde{\bm{k}})=-\bm{g}(\tilde{\bm{k}})$.
$\mu_{\rm B}$ is the Bohr magneton.
We use the unit in which $\hbar=k_{\rm B}=1$.
For the numerical calculation,
we employ the Riccati formalism.
Regarding $\check{K}(\tilde{\bm{k}})$ as the self energy,
one can use the same manner described in Ref.~\cite{higashi2014}.

Starting from the pairing interaction in Ref.~\cite{frigeri2006},
the gap equation for the spin-singlet pair potential $\varDelta(\bm{r})$ is given by \cite{sigrist1991}
\begin{equation}
\varDelta(\bm{r})=\lambda \pi T \cfrac{1}{2} \sum_{-n_{\rm c}(T)-1 < n < n_{\rm c}(T)} \sum_{s^\prime_1 s^\prime_2}
(i \sigma_y)^\dag_{s^\prime_2 s^\prime_1} \left\langle f^0_{s^\prime_1 s^\prime_2}(\bm{r},\tilde{\bm{k}}^\prime,i\omega_n) (i \sigma_y)_{s^\prime_1 s^\prime_2} \right\rangle_{\tilde{\bm{k}}^\prime}.
\end{equation}
$\langle \cdots \rangle_{\tilde{\bm{k}}}$ denotes the average on the circular arc with its radius 1.
The coupling constant is determined from
\begin{equation}
\cfrac{1}{\lambda}=\ln \left(\cfrac{T}{T_{\rm c0}}\right) + \sum_{0 \leq  n < n_{\rm c}(T)}\cfrac{2}{2n+1},
\end{equation}
where $T_{\rm c0}$ is the superconducting transition temperature for $\alpha=t_\perp=\mu_{\rm B}H=0$ and $n_{\rm c}(T)=(\omega_{\rm c}/\pi T-1)/2$.
We fix the cut off frequency to $\omega_{\rm c}=7\pi T_{\rm c0}$.

The local density of states per spin and layer is given by
\begin{equation}
N(E,\bm{r})=-\cfrac{N_{\rm F}}{2N} \cfrac{1}{\pi}\left\langle {\rm Im} \left[{\rm Tr}\hat{g}(\bm{r},\tilde{\bm{k}},i\omega_n \rightarrow E+i\eta) \right] \right\rangle_{\tilde{\bm{k}}},
\end{equation}
where $N_{\rm F}$ is the density of states per spin and layer at the Fermi level in the normal state.

\section{Results and discussions}
\begin{figure}[tb]
\begin{center}
\begin{minipage}{18pc}
\includegraphics[width=18pc]{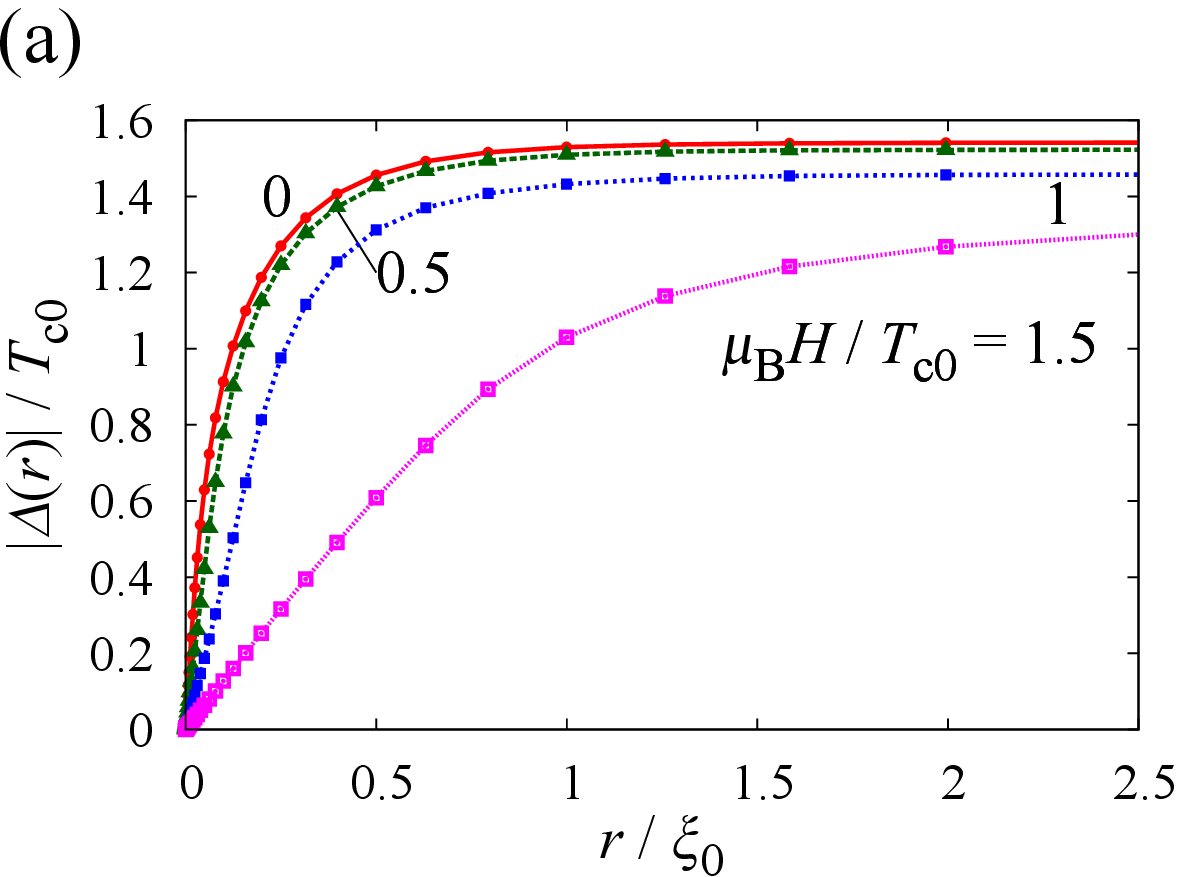}
\end{minipage}\hspace{0.1pc}%
\begin{minipage}{18pc}
\includegraphics[width=18pc]{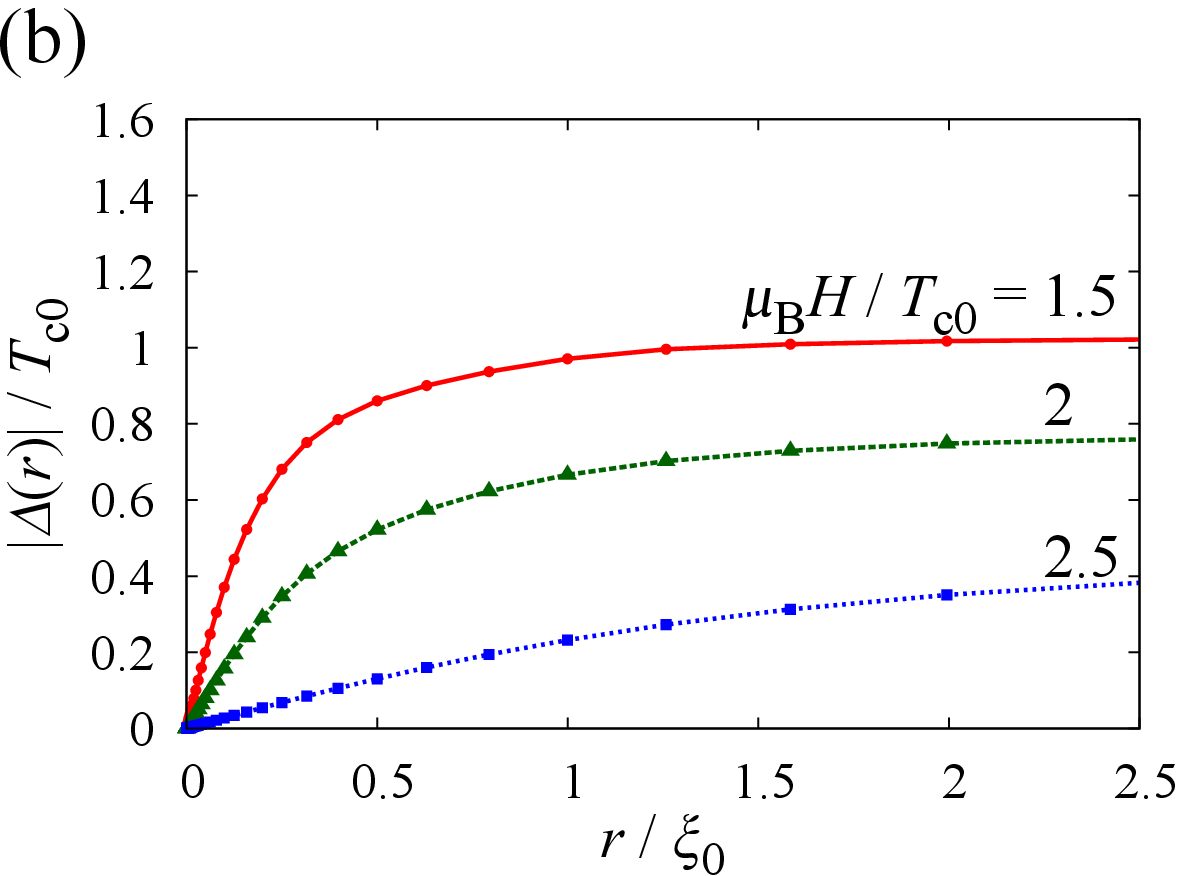}
\end{minipage} 
\caption{
\label{gap}
(color online) Spatial profiles of the pair potential amplitude $|\varDelta(r)|$ for (a) the BCS state and (b) the PDW one.
The horizontal axis represents the radial distance from a vortex center $r=0$.
We set $T/T_{\rm c0}=0.1$, $\alpha/T_{\rm c0}=2$ and $t_\perp/T_{\rm c0}=1$.
The data is plotted for each Zeeman field $\mu_{\rm B}H/T_{\rm c0}$.
}
\end{center}
\begin{center}
\begin{minipage}{18pc}
\includegraphics[width=18pc]{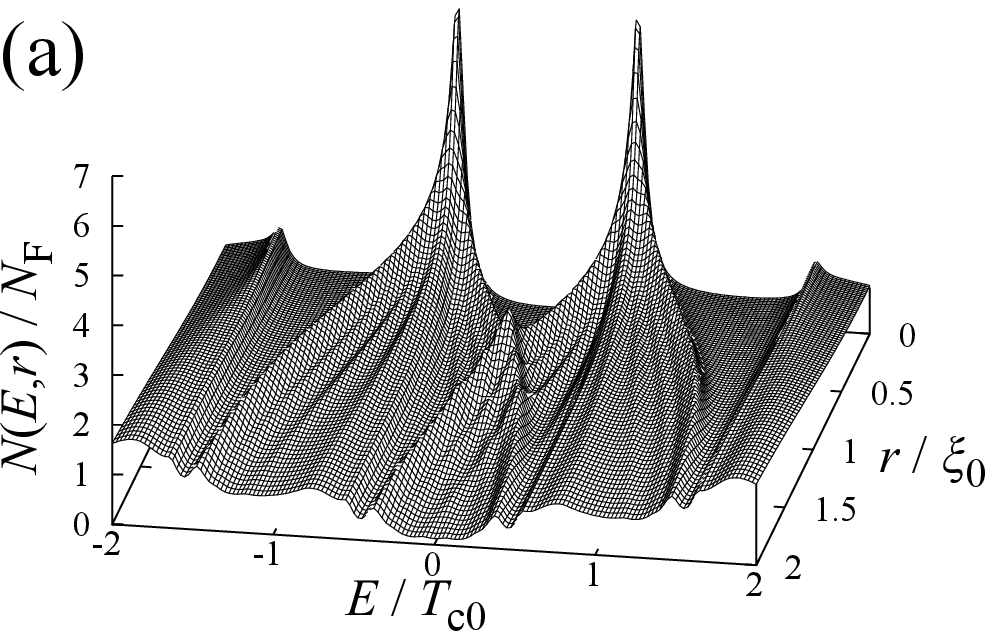}
\end{minipage}\hspace{1pc}%
\begin{minipage}{18pc}
\includegraphics[width=18pc]{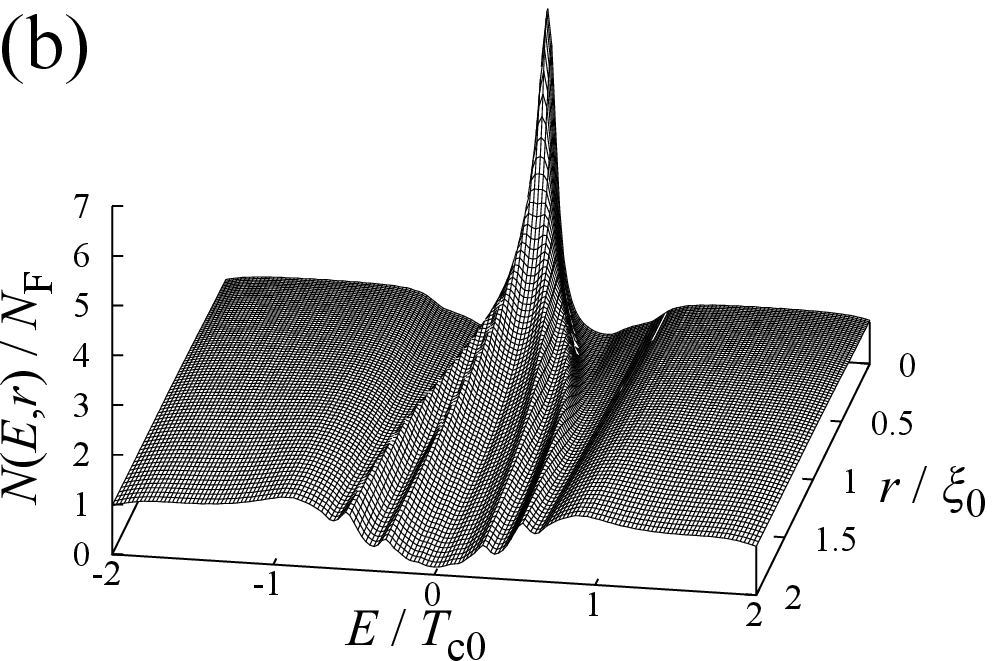}
\end{minipage} 
\caption{
\label{ldos}
The local density of states $N(E,r)$ for (a) the BCS state and (b) the PDW one.
The Zeeman field is set to (a) $\mu_{\rm B}H/T_{\rm c0}=1.5$ and (b) $\mu_{\rm B}H/T_{\rm c0}=2$.
Other conditions are $T/T_{\rm c0}=0.1$, $\alpha/T_{\rm c0}=2$, $t_\perp/T_{\rm c0}=1$ and $\eta=0.05T_{\rm c0}$.
}
\end{center}
\end{figure}

In the numerical results,
we fix the temperature, the SOC and the interlayer hopping to $T/T_{\rm c0}=0.1, \alpha/T_{\rm c0}=2$ and $t_\perp/T_{\rm c0}=1$.
In Fig.~\ref{gap}, we show the spatial variations of the pair potential amplitude around a vortex for (a) the BCS and (b) the PDW states.
The horizontal axis is normalized by the coherence length $\xi_0=v_{\rm  F}/T_{\rm c0}$.
Each plot corresponds to the different magnetic field.
The amplitude of the pair potential gets smaller with increasing the Zeeman field due to the paramagnetic pair breaking effect in both pairing states.
One can notice that the pair potential amplitude of the PDW state is smaller than that of the BCS state.
It is because the PDW state stabilizes in a high magnetic field under the influence of the paramagnetic depairing.
Though the amplitude of the pair potential in the PDW state is smaller than that in the BCS state,
the core size in the PDW state does not spread [compare the profiles for $\mu_{\rm B}H/T_{\rm c0}=1.5$].
The core size in the PDW state for $\mu_{\rm B}H/T_{\rm c0}=1.5$ and $2$ is smaller than that in the BCS state for $\mu_{\rm B}H/T_{\rm c0}=1.5$.
Thus, the sudden decrease in the core radius with increasing the magnetic field is a signature of the BCS-PDW phase transition.
The BCS state is completely destroyed for $\mu_{\rm B}H/T_{\rm c0} \gtrsim 2$,
whereas in the PDW state, the nonzero self-consistent solutions of the pair potential exist even for the high magnetic field $\mu_{\rm B}H/T_{\rm c0} \gtrsim 2$.
The profiles of the pair potential is not altered for $\alpha/T_{\rm c0}=50$.

In the uniform system without vortices,
the stable pairing state prefers the larger bulk amplitude of the pair potential.
In this paper, though there is a vortex in the system,
we consider that the pair potential with the larger bulk amplitude is the more stable solution of the gap equation for a given magnetic field.
The $T$-$H$ phase diagram of bilayer system is displayed in Ref.~\cite{yoshida2012},
which is not altered qualitatively in the vortex state when the system belongs to type-II limit and has the large Maki parameter \cite{yoshida2012,yoshida2014_jpscp}.
According to the field dependence of the pair potential amplitude in the bulk obtained by the quasiclassical theory,
the non-zero self-consistent solution of the gap equation exists only in the PDW state for $\mu_{\rm B}H/T_{\rm c0} \gtrsim 1.8$.
This is consistent with the results calculated from the Bogoliubov-de Gennes equation \cite{yoshida2012}.

In Fig.~\ref{ldos}, we display the energy and spatial dependence of the local density of states (LDOS) around a vortex core.
On the basis of the above consideration,
the calculation of the LDOS is conducted for (a) $\mu_{\rm B}H/T_{\rm c0}=1.5$ in the BCS state
and (b) $\mu_{\rm B}H/T_{\rm c0}=2$ in the PDW one.
As shown in Fig.~\ref{ldos}(a),
the zero energy peak (ZEP) splits off due to the Zeeman field.
The Zeeman split of the ZEP gets suppressed with increasing the SOC strength $\alpha$ for a fixed magnetic field,
since the spin quantization axis is locked within the $x$-$y$ plane by the Rashba SOC.

Contrastingly,
though the PDW state emerges in a high magnetic field, the zero energy peak exists,
which is quite different from the LDOS structure in the BCS state.
If the zero energy quasiparticle states within a core are experimentally observed by scanning tunneling microscopy/spectroscopy (STM/STS),
the STM/STS image of a vortex in the PDW state can be drastically different from that in the BCS state.
To obtain the clue to the LDOS structure in the PDW state,
we start from the case with $\alpha/T_{\rm c0}=0$ for simplicity.
In this case,
the four folding ZEP splits into four peaks.
First, for $\mu_{\rm B}H/T_{\rm c0}=0$,
the band degeneracy is lifted by the interlayer hopping and the ZEP splits into the two peaks.
This split of the ZEP is regarded as the level repulsion of the zero energy bound states.
In the BCS state,
the band degeneracy is not lifted only by the interlayer hopping. 
Then, the $\pi$ phase difference between the layers is essential for the repulsion of the zero energy bound states.
Second, each peak is split off due to the Zeeman field.
Hence the four peaks appear due to the interlayer hopping and the Zeeman field.
Further increasing the magnetic field,
the two peaks in the low energy side shift to the lower energy side and those in the high energy side shift to the higer energy one.
Eventually the two peaks in the low energy side get combined and a single zero energy peak appears.

In the real materials, for example,
the CeCoIn$_5$/YbCoIn$_5$ superlattice,
the superconducting layer is CeCoIn$_5$, which is experimentally identified as a $d$-wave superconductor \cite{sakakibara2007}.
The LDOS structure obtained in this study changes considerably,
since the quasiparticle states around a vortex are sensitive to the pairing symmetry.
The more realistic calculation assuming the $d$-wave pairing symmetry is the future work.

\section{Summary}
We have numerically investigated the electronic structure of a vortex core in bilayer Rashba superconductors by means of the self-consistent quasiclassical calculation.
We found that the local density of states (LDOS) structure in the pair-density wave (PDW) state is quite different from that in the BCS state.
The zero energy vortex bound state exsists in the PDW state, whereas it is absent in the BCS state due to the Zeeman effect.
This characteristic LDOS structure can be observed by scanning tunneling spectroscopy at low temperature.
Another feature of the PDW state is that the core size is small compared with that in the BCS state in the vicinity of the BCS-PDW phase transition.
These features in a magnetic field can be a key to identify the exotic superconducting phase.
\section*{Acknowledgments}
The authors thank M. Kato for helpful discussions and reading this manuscript.
The computation in this work has been done using the facilities of the Supercomputer Center of the Institute for Solid State Physics, the University of Tokyo.
This study has been partially supported by JSPS KAKENHI Grant Numbers 26800197, 24740230 and 25103711.


\end{document}